# Relevance of financial development and fiscal stability in dealing with disasters in Emerging Economies

Valeria Terrones[1] and Richard S.J. Tol[2,3,4,5,6,7]

**Abstract** Previous studies show that natural disasters decelerate economic growth, and more so in countries with lower financial development. We confirm these results with more recent data. We are the first to show that fiscal stability reduces the negative economic impact of natural disasters in poorer countries, and that catastrophe bonds have the same effect in richer countries.

**Keywords** Disasters · Financial development · Fiscal stability · Catastrophe bonds · Disaster risk management · Climate change

**JEL Classification** E17, E62, G00, G20, H30, Q54

[1] Department of Economics, University of Sussex, Falmer, BN1 9 SL United Kingdom
vt205@sussex.ac.uk
[2] Department of Economics, University of Sussex, Falmer, BN1 9 SL United Kingdom
R.Tol@sussex.ac.uk
[3] Institute for Environmental Studies, Vrije Universiteit, Amsterdam, The Netherlands
[4] Department of Spatial Economics, Vrije Universiteit, Amsterdam, The Netherlands
[5] Tinbergen Institute, Amsterdam, The Netherlands
[6] CESifo, Munich, Germany
[7] Payne Institute for Public Policy, Colorado School of Mines, Golden, CO, USA

1. **Introduction**

Empirical evidence suggests that emerging economies are more vulnerable to disasters, the frequency and intensity of which have increased due to the growing population, greater wealth accumulation, and climate change. In 2020, the Centre for Research on the Epidemiology of Disasters (CRED) reported 80 disasters, almost double the amount of 42 disasters in 1980 in emerging economies. The higher frequency and severity of catastrophes imply higher costs for these economies, more so when countries do not have integral risk disaster management. Moreover, in the aftermath of the Covid-19 pandemic, fiscal constraints are raised, especially in those poor countries with previous fiscal imbalances, increasing their vulnerability to deal with any catastrophic event.

The present paper hypothesizes that financial development and fiscal stability play key roles in dealing with disasters in the case of emerging economies. Several studies have researched the relationship between natural disasters, financial development, and economic growth; however, few studies have analysed the impact of fiscal stability prior to the disastrous events that occurred. In this aspect, the literature has focused on what happened after the event and the effect of disasters on government debt. Using a panel fixed effects approach during 1980-2020, and a dynamic panel estimator as a robustness check, we find that financial development and fiscal stability are crucial to dealing with natural disasters in emerging economies.

In addition, the effect of the increasingly popular catastrophic bonds (CAT bonds) is analysed. From their creation in 1997 until 2020, we demonstrate that CAT bonds are a powerful vehicle to deal with natural disasters in advanced economies using a panel fixed effects approach, in contrast with emerging economies where other type of analysis have to be used given the small observations in the data. In fact, more than 80% of the sample corresponds to CAT bonds issued by the United States, Europe, and Japan. More efforts have to be done to improve integrated disaster risk management that incorporates ex-ante and ex-post disaster occurrence.

The paper is structured as follows. Section 2 discusses the literature on the relationship between natural disasters, economic growth, and financial development as well as fiscal stability. Moreover, we introduce the economic theory that justifies the intervention of the government in disaster risk management and the instruments and related policies. Section 3 and 4 describe the secondary data provided by International Monetary Fund and the World Bank; the Centre for Research on the Epidemiology of Disasters at the School of Public Health of the Catholic University of Louvain (Belgium), and a leading alternative risk transfer and weather trading market enterprise, Artemis. Section 5 explains the methodological approach and discusses its main assumptions and limitations. Section 6 presents the main findings of the empirical analysis and the robustness of the analysis. Finally, section 7 summarizes the principal conclusions, adds caveats, and draws policy implications.

## 2. Literature review

2.1 Disasters and economic growth

The literature on the disaster's impact on economic growth remains inconclusive. Some economists argue that natural disasters have a positive effect on economic growth (Albala-Bertrand, 1993; Skidmore and Toya, 2002; Onuma et al. , 2021). Albala-Bertrand (1993) argues that disasters rarely have adverse effects on economic growth because of the regenerative capacity of economies and the disproportionate impact on families from different income levels. Skidmore and Toya (2002) find that climatic disasters are positively correlated with economic growth, human capital investment, and total factor productivity growth in the long run. The authors assert that the reduction of the expected return to physical capital when a disaster occurs has a positive externality, boosting the return on human capital. Additionally, disasters bring the opportunity to introduce new technologies, raising total factor productivity. However, this result is exclusive to climate disasters, not geologic disasters. This is not surprising because the latter type of disaster is more destructive. The process of recovery of physical capital can take years, and human capital can be affected. Despite both papers using all data available at the time, their small samples can lead to estimation errors (28 and 89 countries, respectively).

In contrast, more recent studies show that natural disasters diminish economic growth and are particularly severe in emerging economies, and they can even cause countries to fall into a poverty trap (Raddatz, 2007; Noy, 2009; Tol, 2010, 2019). Using a panel auto-regression approach, Raddatz (2007) quantifies the impact of various external shocks, natural disasters among them, in low-income countries. He finds that climate[8] and humanitarian[9] disasters result in decreases in real per-capita GDP of 2% and 4%, respectively, one year after the event[10]. In the same vein, Noy (2009) finds that the amount of property damage incurred during a disaster is a negative determinant of GDP growth performance using extensive panel data from 1900 to 2008. In particular, developing countries experience higher initial costs and indirect costs on economic activity. Furthermore, this study observes that developing countries, and in particular, small economies, tend to be more vulnerable than larger economies affected by the same events.

The relationship between disasters and economic activity depends on the type and severity of the event as well as particular factors such as the socio-economic and geographic vulnerabilities of each country. These vulnerabilities are increasing as the population raises, people are not aware of the dangers, and they continue to damage the environment, which in turn could intensify the effects of natural disasters. For example, earthquakes are most destructive when steeply sloped land loses grass and forest becomes occupied by informal housing, and droughts are worsened by deforestation, soil erosion, and inadequate land use. Additionally, increasing evidence suggests that this process is being driven by hum greenhouse gas emissions, which may alter weather patterns by raising temperatures and extreme events frequency (Cummins and Mahul, 2008; Franzke, 2017).

---

[8] The definition used by the author includes floods, droughts, extreme temperatures, and windstorms.
[9] It includes famines and epidemics.
[10] Geological disasters have a small and non-significant impact on output. Although this type of events are more destructive, the reason probably is the number of observations in the sample. Only eight countries experienced a geological disaster during 1965-1997.

The foregoing highlights the relevance of understanding what factors influence the appropriate management of dealing with disasters. This research proposes that financial development and fiscal stability play key roles in dealing with disasters, especially in the case of emerging economies. In the short term, the success of the economic recovery process depends on the access and management of economic resources from both the private and public sectors. In developed economies, recovery is paid by insurance, by reserves of households and companies, by the government, or by new loans from commercial lenders. In developing countries, contributions from these sources are scarce, and recovery depends on support from informal networks and the help of the international community (Tol, 2021).

## 2.2 Disasters and financial development

Several studies have demonstrated the association between financial development and economic growth (Noy, 2009; McDermott et al., 2014; Keerthiratne and Tol, 2017). Using panel data on natural disaster events at the country level for the period 1979-2007, McDermott et al. (2014) explore the financial channel -- access to credit in the private sector -- after the event occurs. They find that natural disasters have persistent negative effects on economic growth over the medium to long term when countries have low levels of financial development. Credit constraints in the short-term hamper access to the banking sector for families affected and therefore, undermine the investment in reconstruction. Nevertheless, this analysis is limited to the credit in the private sector which is only one dimension of financial development as is recognized by Keerthiratne and Tol (2017)[11]. In some cases, the public sector is the main source of financing after a disaster. The public sector, in turn, through government programs and public-private partnerships, can provide financial instruments, as we will discuss later.

In the same line, Noy (2009) finds that developing economies and smaller economies face much larger output reductions after a disaster of similar relative magnitude than developed or bigger countries. The author recognizes the important role of appropriate financial conditions to mitigate the negative effects on output. Furthermore, higher literacy rates, stronger institutions, higher per capita income, greater openness to trade, and higher levels of government spending all assist countries in overcoming the disaster in the short term and preventing further economic spillovers. The quality of government and democracy can reduce vulnerability to natural disasters because strong legal and regulatory frameworks might support the development of a sound financial system, particularly insurance services (Tol, 2019). Those characteristics are commonly seen in countries where public finances are stable.

## 2.3 Disasters and fiscal stability

---

[11] The authors find that countries get deeper into debt after a natural disaster, especially in poorer countries whilst the effect is weaker in countries where agriculture is more important.

It is well-recognized that large disasters have fiscal implications for middle- and low-income countries and pose challenges for debt sustainability. Previous studies have emphasized the negative impact on debt after a natural disaster because it can cause an abrupt increase in government spending on relief activities, recovery, and infrastructure reconstruction. Likewise, natural disasters can cause indirect costs as private revenues fall and debt costs rise (Cummins and Mahul, 2008; Borensztein, Cavallo, and Valenzuela, 2009; Koetsier, 2017). Klomp (2017) suggests that a natural disaster increases the probability of a sovereign debt default by around three percentage points, especially in the cases of earthquakes and storms.

In particular, Koetsier (2017) finds that extreme disasters lead to a sizable increase in public debt, but, there is heterogeneity in the magnitudes of these debt accumulations. He employs a panel synthetic control method which constructs a counterfactual for the disaster country for a sample of 163 countries between 1971 and 2014. The author finds that government debt, on average, increases by 11.3% of GDP compared to the synthetic control group. In addition, the median effect on public debt is 6.8% of GDP and some larger disasters can increase public debt by over 20% of GDP. Moreover, Borensztein et al. (2009) show that in the case of Belize, the last two hurricanes caused damage worth 33 percent and 30 percent of GDP, the worst impact registered, and a direct fiscal cost of US$ 50 million. He highlights that debt vulnerability can improve significantly with the use of catastrophic risk insurance.

This paper argues that fiscal stability too is important in dealing with natural disasters in the case of developing countries. Our objective is to understand the contemporaneous relationship between them and the relevance of fiscal stability before the event. Two factors support the relevance of fiscal stability to facing disasters in emerging economies. First, disaster management must be considered from an ex-ante perspective and not only be based on the international donor's assistance. Furthermore, once the event has happened, the interest rates that the country will face to finance the reconstruction will depend, in part, on the payment credibility, transparency, and responsibility of the country's public finances. This is particularly important when countries do not have financial instruments to deal with extreme disasters, which happens in most middle and low-income countries.

### 2.4 Instruments to deal with disasters

Financial instruments to deal with natural disasters come from the private and public sectors, and from public-private partnerships. Cummins and Mahul (2008) suggest that public intervention can be justified in two areas. First, the government can foster an increase in the amount of affordable private insurance coverage purchased by individuals and businesses to protect against catastrophic events with market-oriented regulation and accurate policies to boost competitiveness and efficiency. Property catastrophe risk insurance for households and small businesses, and agricultural insurance are some examples of private financial instruments. Second, sovereign insurance can be created for governments to finance their needs without affecting public fiscal stability or diverting funds from economic development projects (Cummins and Mahul, 2008).

Moreover, a public intervention must be accompanied by a comprehensive disaster risk management strategy[12]. It implies planning to better manage disaster costs; ensure predictable and timely access to funds during the relief, recovery, and reconstruction; and mitigate long-term fiscal impacts. As depicted in Figure 1, rapid access to resources is fundamental in the relief stage. Depending on the probability of disaster occurrence and the severity of its impact, certain financial instruments are more suitable than others, and not all are needed at once. For example, in the relief stage, contingent financing lines and disaster funds are usually recommended to use when disasters are more frequent. As severity rises and frequency declines, some insurance or financing instruments become more relevant such as catastrophe bonds, insurance, and weather/catastrophic derivates in the relief stage (Borensztein, Cavallo, and Jeanne, 2017; World Bank Group, 2021). While immediate liquidity is crucial to support relief and early recovery operations, the government has more time to mobilize resources for the reconstruction.

**Figure 1: Financial instruments by probability and severity of the disaster**

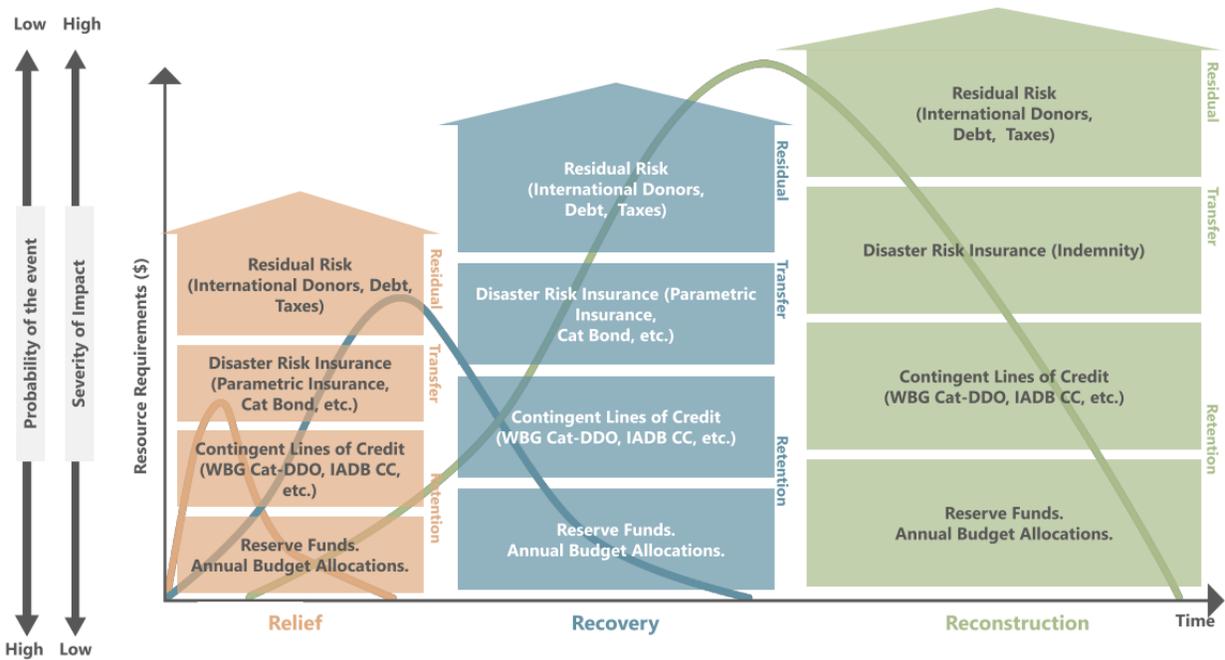

Source: World Bank, 2021.

It is important to note that no single financial instrument can handle all the disaster risks that a country can face. International evidence has shown that governments combine different instruments to protect against events of different frequency and severity (Mahul et. al., 2018). This approach is known as *"risk layering"* which is part of the strategy that mobilizes different financial instruments, before and after a disaster occurs, to address the evolving need for funds (Figure 2). *"Risk layering"* ensures that cheaper monetary resources are used first and that the most expensive tools are used only in exceptional conditions. For example, insurance and CAT bonds can cover extreme events such as earthquakes, but they are not appropriate to protect against low-intensity disasters that occur frequently. The government could consider setting up a contingent fund for those cases of lower layer risk (Figure 2).

---

[12] Mahul, Lukas and Ashok (2018) proposed four core principles of disaster risk finance: timeliness of funding (speed matters but not all resources are needed at the same time), disbursement of funds (how money reaches beneficiaries is as important as where it comes from), disaster risk layering (no single financial instrument can address all disaster risks) and data and analytics (right information is needed to take financial decisions).

**Figure 2: Risk layering approach**

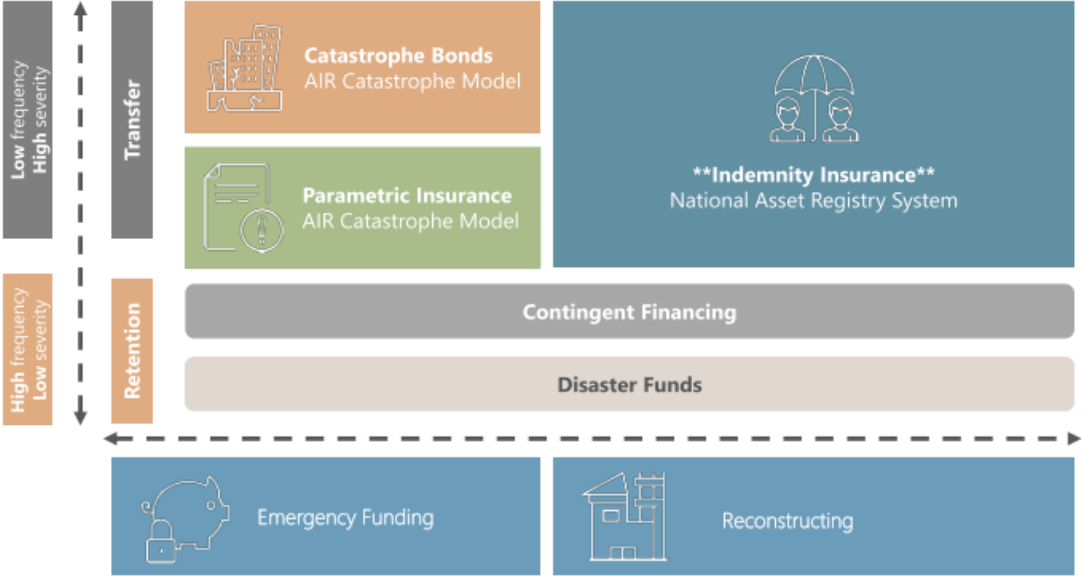

Source: World Bank, 2021.

In particular, in emerging economies, the development of catastrophe insurance and reinsurance markets is affected by demand and supply market imperfections, which justify public intervention (Cummins and Mahul, 2008). On the demand side, low awareness of catastrophe risk exposure, low insurance education and understanding of the benefits[13], weak institutional capacity of governments in disaster risk management, low business volume, and unstable demand for catastrophe insurance are some reasons for the low demand for catastrophe insurance. Insurers may be reluctant to work with the government due to the absence of commitment and long-term vision. In many developing countries, the short-term perspective dominates the political arena. Thus, the incumbent government tends to implement populist policies that help it acquire votes and political backing instead of reforms and far-reaching policies. Much of the work is focused on risk-mitigation investments instead of developing a comprehensive risk-financing strategy.

On the supply side, limited access to capital markets; weak international reinsurance capacity and domestic market capacity; reinsurance cycles[14]; adverse selection, agency costs, and monitoring costs; restricted technical capacity; regulatory impediments; and informational costs are features that influence the development of insurance markets in developing countries (Cummins and Mahul, 2008). Informational asymmetries are especially pronounced when the primary insurer ceding the risk ("the cedent") is a local insurance company, given the limited underwriting data available on insurers and markets. Therefore, adverse selection costs are likely to be relatively high for primary insurers, raising the reinsurance prices. Agency costs arise when managers of insurance companies are out of the common interest of all parties, and consequently, the price of insurance increases. High travel costs are a common example that raises operational costs in the insurance market. Monitoring and disclosure mechanisms to control agency costs are weak in most developing countries. Hence, high expenditure for monitoring and control mechanisms is needed that again raises the cost of insurance.

---

[13] Generally, insurance is perceived as a nonviable financial instrument because indemnities are not paid frequently, but premiums are collected every month and year.
[14] Reinsurance markets experience periodical market fluctuations usually triggered by economic crises.

In contrast with insurance, CAT bonds do not have credit risk and can provide lower spreads because catastrophic events are not correlated with market investment returns (such as equity or commodities). CAT bonds are high layers of financial instruments and have become a standardized tool to manage high-severity and low-frequency disasters such as hurricanes and earthquakes in the last two decades (Cummins and Mahul, 2008). CAT bonds transfer catastrophic risk to the capital markets via the bond issue, avoiding the increase of a country's debt. Repayment of principal is contingent upon the occurrence of a predefined disaster and it is paid out based on a predetermined indemnity level, index, or parametric trigger defined (Cummins, 2012).

Alternative to conventional insurance, CAT bonds cover large amounts of money and have longer periods of coverage, around 3-5 years instead of one-year coverage in the case of insurance (World Bank Group, 2021). Moreover, CAT bonds bring the opportunity to reduce costs and premiums when it is emitted with a pool of countries. For example, Chile, Colombia, Mexico, and Peru (the Pacific Alliance countries) issued a US$ 1,360 billion CAT bond, the largest sovereign risk insurance transaction conducted, and the second largest issuance in the history of the CAT bond market in emerging markets in 2018 (World Bank Group, 2021).

The literature on CAT bonds is recent and there are a few studies that analyses the relationship between this instrument and its effects. Most of the literature is oriented to study the methods to design this instrument to diversify the risk (Coval, Jurek and Stafford, 2009; Deng *et al.*, 2020; Goda, 2021). As it was mentioned above, the CAT bond is an instrument that can be used in advanced and emerging economies following the risk disaster management and the risk layering approach. For the case of developing economies, Borensztein, Cavallo, and Jeanne (2017) find that CAT bonds are an optimal instrument to use against natural disasters (earthquakes, floods, and storms) using a dynamic optimization model. The authors suggest that this instrument may bring a welfare gain of several percentage points of annual consumption by improving the country's ability to borrow, but these gains can disappear when the probability of external debt default is high. In particular, Wolfgang and López (2010) conclude that the use of CAT bonds for earthquakes are an optimal instrument in the case of Mexico. However, there are no papers that study the impacts of CAT bonds on economic growth after a natural disaster. The current paper is the first to do so.

3. **Data**

Data on natural disaster events is obtained from the publicly available Emergency Events Database (EM-DAT) provided by the Centre for Research on the Epidemiology of Disasters at the School of Public Health of the Catholic University of Louvain in Belgium. The data set covers all major disasters across the world from 1900 to the present, compiled from various sources including the United Nations, governmental and non-governmental agencies, insurance companies, research institutions, and press agencies[15].

EM-DAT defines a disaster event as fulfilling at least one of the following criteria: (i) 10 or more people killed; (ii) 100 or more people affected; (iii) a state of emergency declared; or (iv) a request for international assistance. Disaster events can be divided into natural or technological disasters. Our analysis focuses on the former, and we can distinguish six types: geophysical, meteorological, hydrological, climatological, biological, and extra-terrestrial. For example, a geophysical event can be an earthquake, volcanic activity, or mass movement (Table 1).

**Table 1: Types of disasters**

| Types | Sub-types |
|---|---|
| *Geophysical* | *Earthquake, volcanic activity, mass movement.* |
| *Meteorological* | *Storm, extreme temperature, fog.* |
| *Hydrological* | *Flood, landslide, wave action.* |
| *Climatological* | *Drought, glacial lake outburst, wildfire.* |
| *Biological* | *Epidemic, insect infestation, animal accident.* |
| *Extra-terrestrial* | *Impact (e.g., airburst), space weather* |

Source: EM-DAT.

The EM-DAT reports the number of people killed, the number of people affected, and the economic cost of disasters adjusted and non-adjusted by inflation. Considering the number of observations reported and the data reliability, we use the number of people affected. Similar to McDermott et al. (2014), we employ the ratio of the number of people affected by the disaster to the total population for each impacted country. Thus, the ratio takes the value of 1 when the measure surpasses 0.5% and 0 otherwise. Formally, the ratio is as follows:

$$\sum_j \frac{total\ affected_{it,j}}{total\ population_{i,t-1}} > 0.005 \qquad (1)$$

where $j$ is the number of disasters, $i$ is the country and $t$ is the year when the event occurs. The use of a binary disaster variable reduces endogeneity with economic development. Nevertheless, developing countries could be overrepresented because of their greater vulnerability to natural disasters and the possible occurrence of an endogeneity problem. With the objective of addressing this, we use a paned fixed effects approach and a dynamic panel estimator (difference and system Generalised Methods of Moments, or GMM) as a robustness estimator (see Methodology section).

---

[15] CRED, 1998. Available at: *https://www.emdat.be/*

The macroeconomic variables are obtained from the World Development Indicators database, published by the World Bank (WB)[16], and the World Economic Outlook database, published by the International Monetary Fund (IMF)[17]. In the former, the data is available from 1960 to 2020, and in the latter, from 1980 to 2021, and projections are given for the next two years. As we will use both databases, the analysis focus on the period 1980 to 2020 for 196 countries around the world. Gross domestic product per capita (GDPpc), credit as a percentage of GDP, and gross public debt as a percentage of GDP are the main variables used in our analysis. Moreover, we follow the same country classification of the IMF: advanced economies and emerging economies[18].

The credit measure is the value of credits by financial intermediaries to the private sector divided by GDP, as proposed by Levine, Loayza and Thorsten (2009). This measure is well recognized in the literature as a proxy of the level of financial development because it isolates credit issued to the private sector from the credit issued by governments, central banks, and public enterprises. Thus, higher levels of private credit indicate higher levels of financial services and therefore deeper financial intermediary development.

To choose the appropriate fiscal stability variable, we avoid using specific debt thresholds. After the critiques of Reinhard and Rogoff's estimations (2010)[19], many studies highlight the lack of evidence of any particular debt threshold above which medium-term economic growth prospects are compromised by Herndon, Ash and Pollin, (2014). In contrast, public debt trajectory and debt volatility are associated with economic growth. Pescatori, Sandri and Simon (2014) find that countries with high but declining debt appear to grow equally as fast as countries with lower debt, and they find evidence that higher debt is associated with a higher degree of output volatility. For that reason, we employ the change of gross public debt as a percentage of GDP[20] with respect to the previous period as our fiscal stability variable. A higher difference indicates higher volatility and a negative impact on the economic growth rate in the medium term.

Finally, a specific financial instrument, "catastrophe bonds", is included in the analysis. Usually called CAT bonds, these are securities that are not repaid by the issuer if a predefined disaster risk is realized, such as a hurricane or earthquake[21]. Moreover, CAT bonds are issued by three different types of institutions: insurance companies, reinsurers, and government catastrophe funds. The California Earthquake Authority (CEA) and the Florida Hurricane Catastrophe Fund (FHCF) are examples of government catastrophe funds, as is the Fund for Natural Disasters (Fonden) in Mexico.

---

[16] World Bank. Available at: *https://databank.worldbank.org/source/world-development-indicators*
[17] International Monetary Fund. Available at: *https://www.imf.org/en/Publications/WEO/weo-database/2022/April*
[18] International Monetary Fund. Available at: *https://www.imf.org/en/Publications/WEO/weo-database/2022/April/select-aggr-data*
[19] Reinhard and Rogoff argued that public debt/GDP ratios above 90% consistently reduce a country's GDP growth among 20 advanced economies over 1946-2009.
[20] The numerator is the sum of all liabilities that require payments of interest and/or principal by the debtor country to the creditor. This comprises debt liabilities in the form of currency and deposits, debt securities, loans, insurance, pensions and standardized guarantee schemes, and other accounts payable.
[21] Polacek (2018). Available at: *https://www.chicagofed.org/publications/chicago-fed-letter/2018/405*

The data on CAT bonds are obtained from Artemis[22], a leading alternative risk transfer and weather trading market enterprise. They published a comprehensive database containing almost every deal since the first emission in 1997, featuring over 570 catastrophe bond transactions. A bond can cover a specific type of catastrophe (e.g., earthquake) or a pool of catastrophes (multi-peril), and can also be issued for a specific country or a pool of countries. Even a CAT bond can cover a pool of catastrophes for a group of countries, called an international multi-peril deal. However, two limitations of this database have to be highlighted. First, is the lack of information about the maturity of the CAT bonds. They usually have short maturity dates between three-to-five years. For that reason, an average of four years is assumed in the estimation section[23]. Second, there is the possibility that data for Asian countries' deals are not published by Artemis.

## 4. Descriptive statistics

Our sample covers the period from 1980 to 2020 for 196 countries around the world. Table 2 contains summary statistics of the main variables used for our analysis. On average, the GDP per capita grew 1.63% in the 196 countries during the period of analysis. The public credit as a percentage of GDP amounted to 43.8%, on average, and the public gross debt as a percentage of GDP rose to 55.8%, on average, between 1980 and 2020 for the full sample. In addition, on average, the number of people affected by disasters was 1.46 of the total population and its variability was high (std. dev.: 7.9).

**Table 2: Summary statistics**

|  | Mean | Std. Dev. | Min | Max |
|---|---|---|---|---|
| GDP per capita growth (annual %) | 1.63 | 6.11 | -64.99 | 140.37 |
| GDP per capita (constant 2015 US$) | 11203 | 16208 | 167 | 112418 |
| Credit (% of GDP) | 43.75 | 40.83 | 0 | 304.58 |
| Public gross debt (% of GDP) | 55.84 | 45.2 | 0 | 600.12 |
| Total affected by disasters/population | 1.46 | 7.94 | 0 | 40.49 |

Our interest is principally in the growth dynamics in developing countries following extreme events. Table 3 distinguishes between emerging and advanced economies. Emerging economies have been historically more exposed to extreme disasters. Approximately 80% of the total disasters occurred in emerging economies, in contrast to 20% in advanced economies from 1980 to 2020.

Furthermore, the table shows that financial development is markedly lower in emerging economies. In fact, credit in advanced economies is more than three times that in emerging economies. On average, credit was around 31.5% of GDP in emerging economies in the period of analysis (vs. 102.1% in advanced economies). An upward trend in credit was observed in both advanced and emerging economies from 1980 until the International Financial Crisis (2007-2009). Subsequently, credit in advanced economies declined slightly, on average, from 117.9% of GDP in 2010 to 114.7% in 2020, while in emerging economies it continued to rise, from 37.2% to 45.0%, on average, especially because of the increase in China (Annex 2).

---
[22] Artemis. Available at: *https://www.artemis.bm/about/*
[23] Investopedia. Available at: *https://www.investopedia.com/terms/c/catastrophebond.asp*

Regarding government debt, advanced economies had greater public gross debt than emerging economies during the period of analysis (58.7% vs. 54.9%), but the dispersion between countries is higher in emerging economies (39.2% vs. 46.9%). The oil shocks of the 1970s, which forced many oil-importing economies to borrow from commercial banks, and the interest rate increases in industrial countries trying to control inflation, led to an international debt crisis that exploded in Mexico in 1982 (IMF, 2022)[24]. Emerging countries were affected by the rise in international rates and the slump in commodity prices. During the nineties, many emerging countries developed economic reforms based on market-driven policies that helped to stabilize their fiscal position. Nevertheless, a wave of financial crisis swept over East Asia in 1997 (e.g., Thailand, Indonesia, and South Korea).

Since the International Financial Crisis, public debt has grown in both emerging and advanced economies. In advanced economies, the public debt rose from 77.7% of GDP in 2007 to 105.5% of GDP in 2012. Subsequently, it was steadily influenced by the fiscal austerity in the Eurozone until COVID-19, when it jumped from 103.8% of GDP in 2019 to 123.2% of GDP in 2020. In emerging economies, public debt almost doubled from 35.7% of GDP in 2007 to 53.9% in 2019 and soared to 63.9% in 2020. Each economic crisis (International Financial Crisis and COVID-19, in particular) increased government debt by around 20 percentage points interannually in advanced economies (emerging economies: 5 percentage points and 10 percentage points, respectively).

**Table 3: Summary statistics by a group of countries**

**Emerging economies**

|  | Mean | Std. Dev. | Min | Max |
|---|---|---|---|---|
| GDP per capita growth (annual %) | 1.54 | 6.55 | -64.99 | 140.37 |
| GDP per capita (constant 2015 US$) | 5306 | 8420 | 167 | 111574 |
| Credit (% of GDP) | 31.53 | 25.94 | 0 | 182.87 |
| Public gross debt (% of GDP) | 54.94 | 46.91 | 0 | 600.12 |
| Disaster (total affected/population) | 1.80 | 7.94 | 0 | 121.40 |

---

[24] International Monetary Fund. Available at: *https://www.imf.org/external/about/histcomm.htm*

**Advanced economies**

|  | Mean | Std. Dev. | Min | Max |
|---|---|---|---|---|
| GDP per capita growth (annual %) | 1.98 | 3.91 | -54.64 | 24 |
| GDP per capita (constant 2015 US$) | 34262 | 18575 | 4056 | 112418 |
| Credit (% of GDP) | 102.1 | 47.94 | 0.19 | 304.58 |
| Public gross debt (% of GDP) | 58.68 | 39.2 | 0 | 259 |
| Disaster (total affected/population) | 0.13 | 1.52 | 0 | 40.50 |

For complimentary analysis, CAT bonds are studied as a sophisticated instrument to deal with natural disasters between their creation in 1997 and 2020. Most of the issues were observed in the United States (57%), followed by Japan, the European Union, and Australia. Only 8.0% was emitted in the Caribbean countries and other countries in Latin America. CAT bonds are assumed to be cumulative and perpetual. For example, in the year 2020, the variable "CAT bond" assumes the value of the total of all CAT bonds issued in the four years prior to that year. Due to the lack of data availability, the maturity of the CAT bond is assumed to be an average of four years, as, usually, they have short maturity dates between three-to-five years.

## 5. Methodology

A panel fixed effects approach is employed to address the problem of endogeneity in the period 1980-2020, and a dynamic panel estimator as a robustness check on our results. The main specification is presented as follows:

$$\Delta y_{it} = \beta_0 + \beta_1 y_{i,t-2} + \beta_2 D_{it} + \beta_3 credit_{i,t-1} + \beta_4 D_{it} * credit_{i,t-1} + \beta_5 \Delta public\_debt_{it} + \beta_4 D_{it} * \Delta public\_debt_{it} + \theta_i + \theta_t + \varepsilon_{it}$$

where $\Delta y_{it}$ is the annual growth rate of output per capita in the country $i$ for period $t$. The explanatory or independent variables are the ratio of the number of people affected by the disaster divided by the total population in the country $i$ for period $t$ ($D_{it}$). The credit is expressed as a percentage of GDP from the previous period ($credit_{i,t-1}$) and the fiscal stability variable selected ($\Delta public\_debt_{it}$) are included with their respective interactions with the disaster index. Furthermore, the second lag of GDP per capita in logs is incorporated to capture convergence effects between economies.

As it was estimated by McDermott et al. (2014), the interaction between the credit measure and the disaster index ($D_{it} * credit_{i,t-1}$) test directly the effects of financial development on economic growth when severe events occur. We would expect to find a positive sign in the coefficient of the interaction term, suggesting that greater financial development mitigates the growth effects of disasters. Furthermore, the difference in the gross public debt between $t$ and $t-1$ and its interaction with the disaster index ($D_{it} * \Delta public\_debt_{it}$) is added. We would expect to find a negative sign in the coefficient result of the inverse relationship between public debt volatility and economic growth when a disaster happens.

We include country-fixed effects ($\theta_i$) and cluster errors by country to control for any omitted country-specific and time-invariant factors. For example, variables such as culture, country size, and race are time-invariant factors. We also incorporate time-fixed effects ($\theta_t$) to control for any shocks that affect all regions simultaneously such as the business cycle or any economic crisis (e. g. the Covid-19 Pandemic).

As was noticed by McDermott, Barry, and Tol (2014), the principal concern in the estimation strategy is the potential endogeneity of disasters: Natural disasters reduce economic growth, poorer countries are more vulnerable to natural disasters. Even though this concern is partly addressed by the use of a binary index of the disaster occurrence proposed by the authors, the use of the number of total people affected by disasters as the numerator can be debatable. Notwithstanding, the number of people affected is the best variable considering the availability and transparency of the data reported by CRED.

Additionally, the increasingly used and sophisticated financial instrument, a CAT bond, is analyzed in depth. In the same direction as the previous estimation, a panel fixed effects approach is employed between 1997 (the first year of emission) and 2020. The specification is presented as follows:

$$\Delta y_{it} = \beta_0 + \beta_1 y_{i,t-1} + \beta_2 D_{it} + \beta_3 cat\ bond_{i,t} + \beta_4 D_{it} * cat\ bond_{i,t} + \beta_5 credit_{it} + \beta_6 inflation_{it} + \beta_7 government expenditure_{it} + \theta_i + \theta_t + \varepsilon_{it}$$

where $\Delta y_{it}$ is the annual growth rate of output per capita in the country $i$ for period $t$. The new explanatory variables are the issuance of the CAT bond in the country $i$ for period $t$ ( $cat\ bond_{it}$ ) and its interaction when the disaster occurs ($D_{it} * cat\ bond_{it}$). For example, if the country has issued a cat bond in the 4 years before the disaster occurs, the interaction term takes the value of one. The effect of the CAT bond issuance on economic growth is expected to be positive and statistically significant when a disaster occurs for emerging and advanced economies. Moreover, credit as a percentage of GDP, the average inflation rate, and government expenditure as a percentage of GDP are included as control variables.

However, this estimation must be taken with caution since it assumes the fact that CAT bond issuance allows an adequate response to the disaster without considering the time between issuance and disaster. It should be noted that the following regression does not measure the effectiveness of the response to the disaster. More detailed information, such as knowing the specific duration of the CAT bond and whether the contract was exercised, would allow more accurate estimation. Cost-benefit analysis and gap analysis are also recommended to understand this instrument in each particular country. The former calculates the present value of the opportunity cost of the issuance of a CAT bond in contrast to the status quo. The latter compares the potential losses and availability of financing tools to analyse: the reasonableness of the hedge acquisition and its order of priority within a layered strategy, and the probability of exhaustion of each instrument and the whole strategy (Villalobos, 2021).

## 6. Empirical results

### 6.1. Financial development and fiscal stability

Table 4 shows that disasters have a significantly negative effect on economic growth. Similar to McDermott et al. (2014), we find that this relationship is mitigated by the degree of financial market development from 1980 to 2020, as can be seen in the interaction between disaster and credit (model 1). In model 2, the change of the gross public debt as a percentage of GDP is included as a control variable, and it is statistically significant. Hence, less volatility of gross public debt as a proxy of fiscal stability as proposed by Pescatori, Sandri, and Simon (2014), contributes to economic growth.

**Table 4: Economic growth and disasters**

| Variables | Dependent variable: GDP per capita annual growth | | |
|---|---|---|---|
| | Model 1 (Full sample) | Model 2 (Full sample) | Model 3 (Full sample) |
| Lagged GDPpc | -4.722*** | -6.072*** | -6.087*** |
| | (0.737) | (0.796) | (0.799) |
| Disaster | -1.168*** | -1.076*** | -1.142*** |
| | (0.342) | (0.337) | (0.353) |
| Credit | -0.0245*** | -0.0269*** | -0.0273*** |
| | (0.00589) | (0.00808) | (0.00807) |
| Dis*Credit | 0.0192*** | 0.0100* | 0.0121** |
| | (0.00682) | (0.00541) | (0.00582) |
| Public debt | | -0.0743*** | -0.0677*** |
| | | (0.0181) | (0.0202) |
| Dis*Public debt | | | -0.0446 |
| | | | (0.0294) |
| | | | |
| Observations | 4,966 | 3,656 | 3,656 |
| R-squared | 0.159 | 0.265 | 0.267 |
| Number of countries | 193 | 193 | 193 |

Note: Annual data 1980-2020, except where lost due to lags. All models include a constant term and annual fixed effects. Errors clustered at the country level[25]. *t*-statistics in parentheses. *** p<0.01, ** p<0.05, * p<0.1.

Moreover, model 3 tests our hypothesis that fiscal stability may be a significant factor in determining the economic effects of natural disasters for the full sample (Table 4). However, as it is noticed in model 3, the coefficient of the interaction between disaster and public debt is not significant for the full sample. In contrast, our hypothesis holds for the case of emerging countries (Table 5, model 6), suggesting the relevance of fiscal stability to avoid a negative economic impact in those countries. When we disaggregate by the level of income, the interaction term is particularly significant in lower-middle-income countries and low-income countries (Table 6).

The relevance of financial development as well as fiscal stability highlight the importance to boost policies oriented to deepening the financial channel and developing a responsible and transparent public finance, especially in emerging economies where the dependency on the commodities cycle is much higher, and the occurrence of disaster is more frequent and costliest (Table 5, model 6). It begins by shifting the policy perspective from the old, outmoded approach of relying solely on donor community assistance to a comprehensive risk financing approach that must include an ex-ante and ex-post perspective of catastrophic management and where the use of the financial instrument is technically dependent on the probability of occurrence and frequency of the events (Figure 1). According to Cummins and Mahul (2008), Disaster Risk Management (DRM) framework should focus on five-pillar: risk assessment; emergency preparedness; risk mitigation; institutional capacity building; and catastrophe risk financing. Those pillars have to base on the principle that loss of life and economic losses can be reduced by advanced planning and cost-effective investment.

**Table 5: Relevance of fiscal stability in Emerging countries when disasters occurred**

| | Dependent variable: GDP per capita annual growth | | |
|---|---|---|---|
| Variables | Model 4 (Emerging countries) | Model 5 (Emerging countries) | Model 6 (Emerging countries) |

---

[25] As it will noticed in the results, the clustered standards errors only convergence on their true parameter values when the number of clusters (not the number of observations) is large.

| | | | |
|---|---|---|---|
| Lagged GDPpc | -4.673*** | -6.162*** | -6.183*** |
| | (0.762) | (1.236) | (1.238) |
| Disaster | -1.271*** | -1.137*** | -1.223*** |
| | (0.365) | (0.367) | (0.383) |
| Credit | -0.0171* | -0.0207* | -0.0214* |
| | (0.00953) | (0.0121) | (0.0122) |
| Dis*Credit | 0.0245*** | 0.0141** | 0.0168** |
| | (0.00789) | (0.00678) | (0.00711) |
| Public debt | | -0.0716*** | -0.0639*** |
| | | (0.0178) | (0.0200) |
| Dis*Public debt | | | -0.0504* |
| | | | (0.0293) |
| | | | |
| Observations | 4,113 | 2,918 | 2,918 |
| Number of countries | 145 | 145 | 145 |
| R-squared | 0.139 | 0.251 | 0.253 |

Note: Annual data 1980-2020, except where lost due to lags. All models include a constant term and annual fixed effects. Errors clustered at the country level. *t*-statistics in parentheses.
*** $p<0.01$, ** $p<0.05$, * $p<0.1$

**Table 6: Relevance of fiscal stability in the lower middle - and low-income countries**

| Variables | Dependent variable: GDP per capita annual growth | | |
|---|---|---|---|
| | Lower middle and low-income countries | Lower middle-income countries | Low-income countries |
| Lagged GDPpc | -6.831*** | -4.579*** | -13.34*** |
| | (1.303) | (1.169) | (2.802) |
| Disaster | -1.743*** | -1.131*** | -2.450 |
| | (0.510) | (0.387) | (1.519) |
| Credit | -0.00421 | -0.0223 | 0.189*** |
| | (0.0181) | (0.0164) | (0.0619) |
| Dis*Credit | 0.0299** | 0.0183* | 0.0454 |
| | (0.0131) | (0.0101) | (0.0835) |
| Public debt | -0.0287*** | -0.0244 | -0.0230*** |
|  | (0.00753) | (0.0156) | (0.00818) |
| Dis*Public debt | -0.0892*** | -0.102*** | -0.0792*** |
|  | (0.0199) | (0.0297) | (0.0229) |
| Observations | 1,532 | 1,055 | 477 |
| R-squared | 0.256 | 0.326 | 0.264 |
| Number of countries | 77 | 52 | 25 |

Note: Annual data 1980-2020, except where lost due to lags. All models include a constant term and annual fixed effects. Errors clustered at the country level. *t*-statistics in parentheses.
\*\*\* p<0.01, \*\* p<0.05, \* p<0.1

### 6.2. Alternative estimation models

Alternative estimations models are designed to handle the possibility of endogeneity in our key variables. Based on the dynamic panel estimators of Arellano and Bond (1991), Arellano and Bover (1995), and Blundell and Bond, (1998), the generalized method of moments (GMM) is employed to exploit all the linear moment restrictions that follow from the assumption of no serial correlation in the errors, in an equation which contains individual effects, lagged dependent variables and no strictly exogenous variables, as it was initially estimated by McDermott et al. (2014). The findings are presented in table 7, which shows the main fixed effects model. The results on the disaster variable and the interactions with financial development and fiscal stability measures in emerging economies are consistent across each of the estimation methods, giving further confidence in the validity of the estimations.

**Table 7: Alternative estimation models**

|  | Dependent variable: GDP per capita annual growth | | | |
|---|---|---|---|---|
|  | (1) Fixed effects | (2) Fixed effects | (3) Difference GMM | (4) System GMM |
|  | Full sample | Emerging economies | Emerging economies | Emerging economies |
| Lagged GDPpc (in logs) | -6.087*** | -6.183*** | -11.21*** | -8.187*** |
|  | (0.799) | (1.238) | (3.854) | (2.990) |
| Disaster | -1.142*** | -1.223*** | -1.888*** | -1.792*** |
|  | (0.353) | (0.383) | (0.461) | (0.588) |
| Dis*Credit | -0.0273*** | -0.0214* | 0.0662 | 0.153** |
|  | (0.00807) | (0.0122) | (0.0987) | (0.0591) |
| Credit | 0.0121** | 0.0168** | 0.0219** | 0.0209* |
|  | (0.00582) | (0.00711) | (0.00876) | (0.0114) |
| Public debt | -0.0677*** | -0.0639*** | -0.0362** | -0.0407*** |
|  | (0.0202) | (0.0200) | (0.0143) | (0.0150) |
| Dis*Public debt | -0.0446 | -0.0504* | -0.0742*** | -0.0929*** |
| | (0.0294) | (0.0293) | (0.0278) | (0.0250) |
| | | | | |
| Observations | 3,656 | 2,918 | 2,769 | 2,918 |
| Countries | 196 | 145 | | |
| R-squared | 0.267 | 0.253 | | |
| | | | | |
| No. of instruments | | | 95 | 102 |
| Arellano-Bond test AR(1) | | | 0.005 | 0.000 |
| Arellano-Bond test AR(2) | | | 0.170 | 0.110 |
| Hansen test | | | 0.974 | 0.741 |

Note: Annual data 1980-2020, except where lost due to lags. All models include a constant term and annual fixed effects. Errors clustered at the country level. *t*-statistics in parentheses.
*** p<0.01, ** p<0.05, * p<0.1

6.3. Cat bonds: an instrument to deal with natural disasters

The effect of the CAT bond issuance on economic growth is expected to be positive and statistically significant when a disaster occurs for emerging and advanced economies. Nevertheless, Table 8 shows the effect of the use of the CAT bond on economic growth when a disaster occurs is only positive and statistically significant in the case of advanced economies between 1997 and 2020. On average and *ceteris paribus*, the use of the CAT bond increases economic growth by 1.32 percentage points, compared to the status quo of not having a CAT bond when an extreme event occurs. Rapid access to funds – specifically CAT bonds – can improve considerably the response of the economy to the disaster.

**Table 8: Cat bonds analysis in advanced and emerging economies**

| | Dependent variable: economic growth | |
|---|---|---|
| | Developing Economies | Advanced Economies |
| Lagged GDPpc (in logs) | -7.293*** | -10.45*** |
| | (2.099) | (3.571) |
| Disaster | -0.585** | -2.054** |
| | (0.228) | (0.818) |
| Cat bond | 0.378 | -0.129 |
| | (0.622) | (0.569) |
| Dis*cat bond | -0.0215 | 1.317* |
| | (0.754) | (0.771) |
| Credit | -0.0190 | -0.0333* |
| | (0.0174) | (0.0175) |
| Inflation | -0.00678 | -0.297** |
| | (0.00486) | (0.111) |
| Government expenditure | -0.0864 | -0.455*** |
| | (0.0583) | (0.157) |
| | | |
| Observations | 2,782 | 693 |
| Number of countries | 146 | 36 |
| R-squared | 0.210 | 0.585 |

Note: Annual data 1997-2020. All models include constant term and annual fixed effects. Errors clustered at the country level. *t*-statistics in parentheses. *** p<0.01, ** p<0.05, * p<0.1

No significant result is found for emerging economies, probably because of the little use of this instrument during the period of the study. Notwithstanding the increasing use of this financial tool in emerging markets, more than 90% of the CAT bond market is used by advanced economies (United States, Japan, the European Union, and Australia) and only 8.0% of the CAT bonds are issued in the Caribbean countries and other countries in Latin America and Asia. These findings indicate that emerging economies would follow the use of this financial instrument as it was proof of its positive results in advanced economies.

Removing the interaction with the disaster variable, the sign of the tenancy of a CAT bond is positive and statistically significant, for the full sample and for emerging and advanced economies separately (Table 9). Moreover, when we disaggregate the information by the most common types of disaster covered, earthquakes and storms (Table 10), we find a positive effect of the issuance of CAT bonds on economic growth in the full sample. On average and *ceteris paribus*, the issuance of the CAT bond oriented to deal with earthquakes increases economic growth by 0.52 percentage points and 0.68 percentage points in the case of storms, compared to the status quo of not having a CAT bond. These results suggest the relevance of this financial instrument to deal with earthquakes and storms, events that are generally are less frequent and more severe, and it proves the technical accuracy of disaster risk management and the risk layering approach. However, this finding can also indicate a reverse causality, countries that grow faster can afford CAT bonds. As argued in the literature review above, financial development is a key driver of economic growth.

**Table 9: Relevance of holding Cat bonds**

| | Dependent variable: economic growth | | |
|---|---|---|---|
| | Full sample | Developing Economies | Advanced Economies |
| Lagged GDPpc (in logs) | -0.482*** | -0.603*** | -1.576*** |
| | (0.0652) | (0.0864) | (0.227) |
| Disaster | -0.450** | -0.361 | -1.274* |
| | (0.215) | (0.233) | (0.767) |
| **Cat bond** | **0.783*** | **1.171*** | **0.629**** |
| | (0.233) | (0.375) | (0.320) |
| Inflation | -0.0253*** | -0.0239*** | -0.149** |
| | (0.00433) | (0.00461) | (0.0601) |
| Government expenditure | -0.0199*** | -0.0139** | -0.0852*** |

|  | (0.00557) | (0.00627) | (0.0129) |
|---|---|---|---|
| Observations | 4,318 | 3,429 | 887 |
| R-squared | 0.155 | 0.137 | 0.434 |

Note: Annual data 1997-2020. All models include constant term and annual fixed effects. $t$-statistics in parentheses. *** $p<0.01$, ** $p<0.05$, * $p<0.1$

**Table 10: Relevance of holding Cat bonds by type**

|  | Dependent variable: economic growth | | |
|---|---|---|---|
|  | Full sample | Earthquakes | Storms[26] |
| Lagged GDPpc (in logs) | -0.482*** | -0.419*** | -0.439*** |
|  | (0.0652) | (0.0616) | (0.0627) |
| Disaster | -0.450** | -0.464** | -0.459** |
|  | (0.215) | (0.215) | (0.215) |
| Cat bond | 0.783*** | 0.527* | 0.675** |
|  | (0.233) | (0.305) | (0.279) |
| Inflation | -0.0253*** | -0.0254*** | -0.0254*** |
|  | (0.00433) | (0.00434) | (0.00433) |
| Government expenditure | -0.0199*** | -0.0177*** | -0.0188*** |
|  | (0.00557) | (0.00553) | (0.00556) |
| Observations | 4,318 | 4,318 | 4,318 |
| R-squared | 0.155 | 0.154 | 0.154 |

Note: Annual data 1997-2020. All models include constant term and annual fixed effects. Using OLS estimation. $t$-statistics in parentheses. *** $p<0.01$, ** $p<0.05$, * $p<0.1$

## 7. Conclusions

Natural disasters have a significant negative impact on economic growth. On average and *ceteris paribus*, a disaster reduces the growth rate of gross domestic economic per capita by 1.17 percentage points. Financial development and fiscal stability are key factors in reducing the negative impact of natural disasters in emerging economies. CAT bonds mitigate the impact on economic growth when a disaster occurs in advanced economies. In emerging economies, there is no significant result, probably due to the small number of observations.

The results confirm that natural disasters are bad for economic growth, and more so in countries that are financially less developed. We show, for the first time, that financial stability ameliorates the economic impact of natural disasters, in poorer countries, and so do CAT bonds, in richer countries. The following caveats apply. Any analysis with the CRED data suffers from reverse causality, as only extreme events that do damage are reported. The discretization by McDermott, Barry, and Tol (2014) partly but not necessarily completely overcomes this problem. The data have a low temporal and spatial resolution; a natural disaster in January is assumed to have the same impact as a natural disaster in December, and a disaster that heavily affects a small part of the country is recorded the same as a disaster that lightly affects the whole country. The indicators for financial development and fiscal stability used here are both common and appropriate, but certainly not the only choice. Finally, there is a room for improvement in disaster financial instruments data. A database of CAT bonds is created by Artemis, unfortunately lacking data on maturity. Data for other instruments are limited. Efforts can be oriented to enhance and compile data on different financial instruments. It would permit us to compare and analyze historically the effectiveness cost of different instruments between countries across time. Future research should address these issues.

---

[26] Cyclones, hurricanes, typhoon, windstorm, thunderstorm, and wind were classified as "storms".

The policy implications are as follows. The results underline that natural disasters have a lasting effect on economic development. Financial development is important for a number of reasons, but also to mitigate the impact of natural disasters. The same is true for fiscal stability. Governments in disaster-prone areas should work to improve both financial development and fiscal stability. CAT bonds are an effective instrument to reduce the negative economic impact of natural disasters and should be deployed more widely.

## 8. References

Albala-Bertrand, J.M. (1993) 'The Political Economy of Large Natural Disasters with Special Reference to Developing Countries', 26(1), pp. 272–273.

Artemis (1999) *Artemis*. Available at: https://www.artemis.bm/about/.

Blundell, R. and Bond, S. (1998) 'Initial conditions and moment restrictions in dynamic panel data models', *Journal of Econometrics*, 87(1), pp. 115–143. Available at: https://doi.org/10.1016/S0304-4076(98)00009-8.

Borensztein, E., Cavallo, E. and Jeanne, O. (2017) 'The welfare gains from macro-insurance against natural disasters', *Journal of Development Economics*, 124, pp. 142–156. Available at: https://doi.org/10.1016/j.jdeveco.2016.08.004.

Borensztein, E., Cavallo, E. and Valenzuela, P. (2009) 'Debt Sustainability Under Catastrophic Risk: The Case for Government Budget Insurance', *Risk Management and Insurance Review*, 12(2), pp. 273–294.

Coval, J.D., Jurek, J.W. and Stafford, E. (2009) 'Economic catastrophe bonds', *American Economic Review*, 99(3), pp. 628–666. Available at: https://doi.org/10.1257/aer.99.3.628.

CRED, C. for research on the E. of D. (1998) *The International Disaster Database*. Available at: https://www.emdat.be/ (Accessed: 30 May 2022).

Cummins, J.D. (2012) 'Cat Bonds and Other Risk-Linked Securities: Product Design and Evolution of the Market', *SSRN Electronic Journal*, pp. 1–22. Available at: https://doi.org/10.2139/ssrn.1997467.

Cummins, J.D. and Mahul, O. (2008) *Catastrophe Risk Financing in Developing Countries*, *Catastrophe Risk Financing in Developing Countries*. Available at: https://doi.org/10.1596/978-0-8213-7736-9.

Deng, G. *et al.* (2020) 'Research on the Pricing of Global Drought Catastrophe Bonds', *Mathematical Problems in Engineering*, 2020. Available at: https://doi.org/10.1155/2020/3898191.

Franzke, C.L.E. (2017) 'Impacts of a Changing Climate on Economic Damages and Insurance', *Economics of Disasters and Climate Change*, 1, pp. 95–110.

Goda, K. (2021) 'Multi-hazard parametric catastrophe bond trigger design for subduction earthquakes and tsunamis', *Earthquake Spectra*, 37(3), pp. 1827–1848. Available at: https://doi.org/10.1177/8755293020981974.

Herndon, T., Ash, M. and Pollin, R. (2014) 'Does high public debt consistently stifle economic growth? A critique of Reinhart and Rogoff', *Cambridge Journal of Economics*, 38(2), pp. 257–279. Available at: https://doi.org/10.1093/cje/bet075.

International Monetary Fund (2022) *World Economic Outlook Database*. Available at: https://www.imf.org/en/Publications/WEO/weo-database/2022/April.

Investopedia (2020) *Catastrophe Bond (CAT) Meaning, Benefits, Risk, Example*. Available at: https://www.investopedia.com/terms/c/catastrophebond.asp (Accessed: 25 June 2022).

Keerthiratne, S. and Tol, R.S.J. (2017) 'Impact of Natural Disasters on Financial Development', *Economics of Disasters and Climate Change*, 1(1), pp. 33–54. Available at: https://doi.org/10.1007/s41885-017-0002-5.

Klomp, J. (2017) 'Flooded with debt', *Journal of International Money and Finance*, 73, pp. 93–103. Available at: https://doi.org/10.1016/j.jimonfin.2017.01.006.

Koetsier, I. (2017) 'The fiscal impact of natural disasters', *U.S.E Discussion Paper Series*, 17(17). Available at: www.uu.nl/rebo/economie/discussionpapers.

Levine, R., Loayza, N. and Thorsten, B. (2009) 'Financial intermediation and growth: Causality and causes without outliers', *Portuguese Economic Journal*, 8(1), pp. 15–22. Available at: https://doi.org/10.1007/s10258-009-0035-y.

Mahul, Oliver; Lukas, Benedikt; Ashok, K. (2018) *Disaster Risk Finance: A Primer Core Principles and Operational Framework*. Washington D.C. Available at: https://www.financialprotectionforum.org/publication/disaster-risk-finance-a-primercore-principles-and-operational-framework.

McDermott, T.K.J., Barry, F. and Tol, R.S.J. (2014) 'Disasters and development: Natural disasters, credit constraints, and economic growth', *Oxford Economic Papers*, 66(3), pp. 750–773. Available at: https://doi.org/10.1093/oep/gpt034.

Noy, I. (2009) 'The macroeconomic consequences of disasters', *Journal of Development Economics*, 88(2), pp. 221–231. Available at: https://doi.org/10.1016/j.jdeveco.2008.02.005.

Onuma, Hiroki; Shin, Kong Joo; Managi, S. (2021) 'Short-, Medium-, and Long-Term Growth Impacts of Catastrophic and Non-catastrophic Natural Disasters', *Economics of Disasters and Climate Change*, Vol. 5 (1), pp. 53–70.

Pescatori, A., Sandri, D. and Simon, J. (2014) 'Debt and Growth: Is There a Magic Threshold? IMF Working Paper Research Department Debt and Growth: Is There a Magic Threshold?', pp. 1–19.

Polacek, A. (2018) *Catastrophe Bonds: A Primer and Retrospective*, *Federal Reserve Bank of Chicago*. Available at: https://www.chicagofed.org/publications/chicago-fed-letter/2018/405.

Raddatz, C. (2007) 'Are external shocks responsible for the instability of output in low-income countries?', *Journal of Development Economics*, 84(1), pp. 155–187. Available at: https://doi.org/10.1016/j.jdeveco.2006.11.001.

Skidmore, M. and Toya, H. (2002) 'Do natural disasters promote long-run growth?', *Economic Inquiry*, 40(4), pp. 664–687. Available at: https://doi.org/10.1093/ei/40.4.664.

Tol, R.S.J. (2010) 'The economic impact of climate change', *Perspektiven der Wirtschaftspolitik*, 11(SUPPL. 1), pp. 13–37. Available at: https://doi.org/10.1111/j.1468-2516.2010.00326.x.

Tol, R.S.J. (2019) *Climate Economics: Economic Analysis of Climate, Climate Change and Climate Policy*. Available at: https://sites.google.com/site/climateconomics/home/about-the-author?authuser=0.

Tol, R.S.J. (2021) 'State capacity and vulnerability to natural disasters', *Working Paper Series 0721* [Preprint]. Available at: http://arxiv.org/abs/2104.13425.

Villalobos, J.A. (2021) 'Development of Catastrophe Bonds for Sovereign Disaster Risk Transfer', in. Washington D.C.

Wolfgang, K. and López, B. (2010) 'Calibrating CAT Bonds for Mexican Earthquakes', *The Journal of Risk and Insurance*, 77(3), pp. 625–650.

World Bank (2022) *World Development Indicators*. Available at: https://databank.worldbank.org/source/world-development-indicators.

World Bank Group (2021) 'From COVID-19 Crisis Response to Resilient Recovery - Saving Lives and Livelihoods while Supporting Green, Resilient and Inclusive Development (GRID)', *Development Committee Meeting*, pp. 1–31.